\documentclass[12pt]{article}
\pdfoutput=1

\usepackage{jheppub}

\usepackage{amsmath}
\usepackage{amssymb}
\usepackage{graphicx}
\newcommand{\be}{\begin{equation}}
\newcommand{\ee}{\end{equation}}
\newcommand{\bea}{\begin{eqnarray}}
\newcommand{\eea}{\end{eqnarray}}
\newcommand{\nn}{\nonumber}

\title{
\boldmath QGP time formation in  holographic shock waves  model of  heavy ion collisions\footnote{Based on invited talks at International Conference on "Quark Confinement and the Hadron Spectrum XI" (confinement XI), St. Petersburg, Russia, 7-13 Sep. 2014
and International Conference on Physics
"In Search of Fundamental Symmetries"â
dedicated to the 90-th birthday anniversary of
          Yu.V. Novozhilov, 1-5 Dec. 2014.}}
\author[]{Irina Ya. Aref'eva}
\affiliation[]{Steklov Mathematical Institute, Russian Academy of Sciences,\\Gubkina str. 8, 119991, Moscow, Russia}
\emailAdd{arefeva@mi.ras.ru}

\abstract{We estimate the thermalization time in two colliding shock waves
holographic model of heavy-ion collisions. For this purpose  we model the process by  the Vaidya metric with a horizon defined by the trapped surface location. We consider two bottom-up AdS/QCD models that give, within the colliding shock waves approach,   the dependence of multiplicity on the energy compatible with RHIC and LHC results. One model is a bottom-up AdS/QCD confining model and the other is related to an anisotropic thermalization.
We estimate the thermalization time and show that increasing the confining potential  decreases   the thermalization time as well as an  anisotropy  accelerates the thermalization.}


\begin{document}
\maketitle


\newpage


\section{Introduction}

Holographic duality \cite{Malda,GKP,Witten} provides a powerful tool for studying static properties of the QGP as well as its thermalization
 \cite{1101.0618,IA,DeWolf}.
There are holographic models that reproduce perfectly the static properties of the QGP, meanwhile others holographic models are used to get non-static characteristics such as the thermalization time
 in heavy-ion collisions and  the charged multiplicity. Holographic thermalization means a black hole formation in the dual
space-time
and particle multiplicities is defined by the entropy of the produced black hole.

The gauge/string duality \cite{Malda,GKP,Witten} perfectly works for the  $\mathcal{N}$=4 Supersymmetric   Yang-Mills theory, while a true  dual description of real  QCD is unknown, in spite
of a lot of efforts have been  made  to find holographic QCD from string setup
\cite{0003136,0209211,0306018,Mateos,SS,EKSS2005,Karch2006}.  This approach is known as the "top-down" approach.
  Other approach, known  as the
"bottom-up" approach,
is supposed to  propose a  holographic
QCD model, i.e. a suitable vacuum background, that fits  experimental data and lattice results.
The quark confining backgrounds that reproduce the Cornell potential,  $\rho$-meson spectrum etc. have been proposed in \cite{Andreev:2006ct,Andreev:2006cu,White:2007tu,Pirner:2009gr,0911.0627,He:2010ye}. Improved holographic QCD (IHQCD) that  reproduce  the  QCD $\beta$-function
have been proposed in \cite{Gursoy:2008za,Gursoy:2010fj}.
The dual holographic approach  has been successfully used to describe the static properties of  the QGP, see \cite{1101.0618} for review.

 The problem of the QGP formation in HIC is the subject of intensive study within the holographic approach in last years (see reviews
 \cite{IA,DeWolf,Chesler:2015lsa} and refs therein) and
 initially  has been considered in the AdS background  \cite{Gubser}-\cite{ABP}, that
 cannot describe either   quark confinement either  reproduce the QCD $\beta$-function.
 However, as we  have  just mentioned, there are backgrounds which solve one, or even two of these problems.
Therefore, to describe the holographic thermalization in more realistic frameworks, it is natural to study thermalization in these backgrounds.

Thermalization in the improved holographic background   has been studied in \cite{KT}  and it has been shown
 that  without additional assumptions, such as  an energy dependent cut-off at the high energy \cite{Gubser2},  one
cannot reproduce the multiplicity dependence on energy observed at RHIC and LHC
in this background.
In  \cite{APP}  it has been noticed that the holographic realization of the  experimental
 dependence of multiplicity on the energy \cite{ATLAS:2011ag} requires an unstable background.
In \cite{Ageev:2014mma} it has been shown that  the  model  reproducing the Cornell potential
also gives an observed energy dependence of multiplicities if one assumes that the multiplicity is related to the
dual entropy produced  during a limited time period.
However in this consideration there is a limitation on the possible energy of colliding shock walls \cite{Ageev:2014mma}. Since in this consideration we have used a more or less general background
reproducing AdS at UV and confinement at IR, we can think  that just  general  assumptions
about the background prevent to reproduce suitable behavior in UV and IR and in the same time give the correct energy dependence of multiplicity at {\it high energy}. In particular, we can think that a default assumption of the
 isotropic form of  background metrics is responsible for this discrepancy
 and   for a more realistic description of  entropy production in holographic models one
has to consider  anisotropic backgrounds \cite{Aref'eva:2014AG}.

In favor of anisotropic holographic backgrounds  there are also additional arguments (see review \cite{Giataganas:2013lga} for holographic studies of strong coupled anisotropic theories).
Up to a year ego, it was believed  that just after heavy-ion collisions, a pre-equilibrium
period exists for up to 1 fm/c and then  the QGP appears and this QGP is
isotropic.  However   now there  is a belief that    the QGP  is created after a very short time after the collision,
$\tau _{therm}\sim 0.1 fm/c$,
and it is anisotropic ("anisotropic" means to a spatially anisotropy)  for  a short time $\tau$ after the collision,
$0<\tau_{therm}<\tau <\tau_{iso}$, and the time of locally isotropization is about
$\tau_{iso}\sim 2 fm/c$ \cite{Strickland:2013uga}. In the holographic version of this setup it is suitable to consider a black hole formation in a spatially anisotropic background.
Motivated by recent experimental indications in favor of anisotropic thermalization,
we also discuss a holographic thermalization scenario  in the anisotropic 5-dimensional Lifshitz-like
background.  The collision of  domain walls in this background has been recently considered in \cite{Aref'eva:2014AG}, where it has been shown
 that for  the critical exponent specifying the Lifshitz-like background equal to 4,
the dependence of multiplicity on the energy  is desirable  $E^{1/3}$.

In this paper we estimate thermalization time for two colliding shock waves in different backgrounds. Our main idea is very simple -- the black hole creation in two shock waves collisions is modeled by  Vaidya metric with a horizon corresponding to the location of the trapped surface appearing in two shock waves collision and thermalization  time
 is estimated within standard prescription in  the Vaidya metric \cite{Keski}. The Vaidya deformation of  isotropic backgrounds   have been successfully used in description of thermalization in several isotropic models \cite{Keski}-\cite{1305.3267} as well as Vaidya deformations \cite{Keranen:2011xs,alishahiha:2012,1401.2807,Fonda:2014ula} of anisotropic  metrics \cite{Taylor:2008tg,Tarrio:2011de,Gouteraux:2011ce,Huijse:2011ef,Dong:2012se,
 Bueno:2012sd,1212.2625}.

As mentioned above,  only for a special background the entropy of the  black hole produced in the
domain shock wave  collision reproduces the energy dependence of  particle multiplicities
obtained at RHIC and LHC.  We estimate   the thermalization time
for these cases. Namely, we  estimate the isotropic/anisotropic thermalization time  in
a holographic bottom-up AdS/QCD  confinement background  that
provides the Cornell potential and  QCD $\beta$-function.
 Here we use the Vaidya deformation of confinement  background metrics in both isotropic and anisotropic cases. We also compare   these  results with our
  previous results  obtained by  general causality arguments \cite{Ageev:2014mma}

The paper is organized as follows.
In Section 2 we discuss the isotropic case and in Section 3 the anisotropic one. Both sections  are started by  setup, where notations and review of previous results are presented.
In main parts of  the sections estimations of thermalization time by the Vaidya modeling are performed.
A comparison of results obtained by Vaidya modeling and  by general causal arguments
are presented in the end of these sections.

\section{Thermalization in isotropic backgrounds}
\subsection{Setup}
\subsubsection{General isotropic metric}
In a general isotropic  holographic approach, the 5-dimensional metric is
\be
\label{metric-b}
ds^2=b^2(z)(-dt^2+dz^2+dx_i^2),\,\,\,\,i=1,2,3.
\ee
Following \cite{KT}  we consider the following form of the b-factor
\be
\label{bz}
b(z)=\frac{e^{cz^2/4}}{z^a},\ee
where $a$ and $c$ are some constants. Metric (\ref{metric-b}) in the top-down approach supposes to solve
 the 5-dimensional dilaton-gravity equations of motion \cite{Gursoy:2008za,Gursoy:2010fj}.
  It is also considered in the bottom-up approach, in particular, the  confining metric considered in  \cite{Andreev:2006ct} corresponds to $a=1$ and $c=0.42\,$GeV$^2$, see also \cite{White:2007tu,Pirner:2009gr,0911.0627,He:2010ye}.

\subsubsection{Thermalization due to shock waves collision}

{\it Point-like shock waves}
are usually considered in top-down backgrounds and they are supposed to be solutions of the 5-dimensional (dilaton) gravity  with point-like sources.
In the case of the point-like shock wave the deformation of metric (\ref{metric-b}) has the form
\bea\nn
ds^2_{shock}&=&ds^2+\delta ds_{shock}^2,\\
\label{shock}
\delta ds^2_{shock}&=&b^2(z)\,\phi(z,x_1,x_2)\delta (u)du^2.\eea
Here and below $u,v=t\pm x_3$ and the point shock profile $\phi(z,x_1,x_2)$ solves the equation
\be
\square_{3b}\phi(z,x_1,x_2)=-16 \pi G_5\,J_{uu},\,\,\,\,\,\square_{3b}=(\partial _\alpha \partial _\alpha +3\frac{\partial _z b}{b}+\partial^2 _z), \ee
here $\square_{3b}$ is the Beltrami-Laplace operator corresponding to metric
$ds^2_{3b}=b^2(z)(dz^2+dx_\perp^2)$, $\perp=1,2$,
$G_5$ is the 5-dimensional gravitational constant and $J_{uu}$ is the component of the energy-momentum tensor sourcing the point-like shock wave, $J_{uu}\sim \delta(u)\delta(z-L)\delta(x_\perp)$.

Two colliding point-like shock waves
\be
\label{2-shock}
\delta ds_{2-shocks}^2=b^2(z)\,(\phi(z,x_1,x_2)\delta (u)du^2
+\phi(z,x_1,x_2)\delta (v)dv^2),\,\,\,\,\,\,u,v<0,\ee
produce a black hole, whose  entropy ${\cal S}$ can be estimated   by the area of the trapped surface
\be
\label{A-is}
{\cal S}\geq {\cal S}_{TS},\,\,\,\,\, {\cal S}_{TS}=\frac{2}{4G_5}\int _{z_a}^{z_a}\sqrt{\det |g _{3b}|}dz dx_\perp^2=
\frac{\pi}{2G_5}\int _{z_a}^{z_b}b(z)^3 x_\perp^2 (z)dz. \ee
  ${z_a}$ and ${z_b}$ are the points where  $\psi (z_{a,b})=0$. $\psi (z)$ is the  trapped surface profile function, which for the  central collision \cite{0805.2927,KT} up to the boundary conditions (that in fact define $z_{a,b}$) satisfies  the same equation as the shock wave profile.

The  simplest form of the black hole
in the background  metric (\ref{metric-b}) is given by \cite{Gursoy:2008za}
\bea
\label{metric-bh}
ds^2&=&b^2(z)(-f(z_h,z)dt^2+\frac{dz^2}{f(z_h,z)z^2}+dx_i^2),\,\,\,\,i=1,2,3,
\\ \label{metric-f}
f(z_h,z)&=&1-K(z_h,z),\,\,\,\,\,K(z_h,z)=\frac{K(z)}{K(z_h)},\\
\label{Kz}K(z)&=&\int _0^z\frac{dz}{b(z)^3},\eea
here $z_h$ is the position of the horizon, and the temperature and  entropy are
\bea
\label{T}
\frac{1}{T}&=&\frac{4\pi}{f^\prime(z_h)}=4\pi\int _0^{z_h}\frac{b(z_h)^3}{b(z)^3}dz\,,\\
\label{A}
{\cal S}_f&=&\frac{b^3(z_h){\cal V}_3}{4G_5}\,,\eea
where ${\cal V}_3$ is the volume of the 3-dim space.

For the {\it shock domain walls} \cite{Lin,ABP}
 the  wave profile does not depend on the transversal coordinates
 and solves
\be
\label{eq31}
\left(\partial^2_z+\frac{3b^\prime}{b}\partial_z\right)\phi^w(z)=
-\frac{16\pi G_5 E}{L^2}\frac{\delta(z-z_*)}{b^3(z)}.\ee
The trapped surface is  located between points $z_a$ and $z_b$
which satisfy equations \cite{APP}
 \bea
\label{rel-E}\frac{8\pi G_5 E}{L^2}b^{-3}(z_a)\int_{z_b}^{z_*} b^{-3}dz&=&\int^{z_a}_{z_b} b^{-3}dz,\\\nn
\frac{8\pi G_5 E}{L^2}b^{-3}(z_b)\int_{z_a}^{z_*} b^{-3}dz&=&-\int_{z_b}^{z_a} b^{-3}dz,
\eea
here $z_*$ is the position of the collision point. The  area density $s$ (per the area in the transversal direction) of the trapped surface located between points $z_a$ and $z_b$, is given by

 \bea\label{gen-entr}
 s=\frac{1}{2G_5}\int^{z_b}_{z_a}b^{3}\,dz.\eea

   Equations (\ref{rel-E})  give the relation between points $z_b$, $z_a$ and the energy $E$,
 \be
 \label{a-b-E}
 b^3(z_a)+b^3(z_b)=\frac{8\pi G_5 E}{L^2}.
 \ee

\newpage
\subsection{Thermalization Times}
  \subsubsection{Estimation with the Vaidya metric }

 In this section we model the black hole creation in two shock waves collision by the  Vaidya metric with a horizon corresponding to the location of the trapped surface appearing in these shock waves collision and estimate the thermalization  time
 within the  standard prescription in  the Vaidya metric \cite{Keski}.

 We relate $z_a$ and  $z_b$, defining the location of the trapped surface, with  the masses $M_{a}$ and $M_{b}$ of the black brane in the background (\ref{metric-b}),
 \be
 \label{Ma}
  M_a\equiv M(z_a),\,\,\,\,\,M(z_a)=K^{-1}(z_a),\ee
  where $K(z)$ is defined by (\ref{Kz}),
 and the same for $z_b$.  We assume that from $z_a<z_b$ it follows that $M_b<M_a$.
The appearance of the trapped surface located at $z_a$ and $z_b=\infty$ can be  modeled by the Vaidya metric
  \bea
\label{ds-f-b}
ds^2&=&b^2(z)\left(-f(z_h, z,v)\,dv^2-2dvdz+d\vec{x}^2\right),\\
f(z_a, z,v) &=& 1-\theta (v)\,K(z_a,z).\eea

As it is accepted in the Vaidya approach \cite{Keski}-\cite{1305.3267}, to find the thermalization time $\tau$ at the scale $\ell$ one has to consider a geodesic  with  equal time endpoints  located at the boundary at  distance  $\ell$ and find the time
$\tau$,  when this geodesic is covered by the black shell (\ref{ds-f-b}). The  thermalization time $\tau$
and the  distance $\ell$ are related as
 \bea
 \label{xM}
\ell&=&
 2s\int_0^{1}\frac{ b(s)}{b(sw)}\,\frac{dw}{\sqrt {\left(1-K(z_h,sw)\right) \cdot\left(1-\frac{b^2(s)}{b^2(sw)}\right)}};\\
 \label{tM}
 \tau &=&
s\int _{0}^{1}\frac{dw}{1-K(z_h,sw)},
\eea
i.e. one finds  the thermalization time at the scale $l$   excluding the axillary parameter $s$
from the system of equations (\ref{xM}) and (\ref{tM}),
\be
 \tau  =\tau_{therm}(z_h, l).\ee
Assuming  $z_a<z_b$   we estimate the thermalization time  due to  the trapped surface formation
  by
 \be
 \label{therm-zazb}
 \tau_{therm}(z_a,z_b,l)=\tau_{therm}(z_a,l).\ee

 \subsubsection{Estimation for confining metric}\label{S:2.2.2}
In this section we consider thermalization for the metric with the confining
 $b$-factor
\be
\label{AZ}
 b(z)=\frac{e^{cz^2}}{z},
 \ee
here $c$ is related with the notations of \cite{Andreev:2006cu}
as $
c=\frac{1}{4}c_{AZ}$ and we  use the bottom-up version of the blackening function considered in \cite{Andreev:2006ct}
\be
\label{M-AZ}
K(z)=z^4,\,\,\,\,f(z_h,z)=1-\frac{z^4}{z_h^4}.\ee
Note that according (\ref{Ma})
there are  non-leading corrections to  (\ref{M-AZ}). Indeed, for the $b$-factor (\ref{AZ})
the blackening function according (\ref{Ma}) is
 \be
 \label{MaAZ}
 K(z_a,z)=\int _0^{z_a}\frac{dz}{b(z)^3}=
  {\frac {-1+{{\rm e}^{-3\,c{z}^{2}}}+3\,{{\rm e}^{-3\,c{z}^{2}}}c{z}^{2
}}{-1+{{\rm e}^{-3\,c{z_{{h}}}^{2}}}+3\,{{\rm e}^{-3\,c{z_{{h}}}^{2}}}
c{z_{{h}}}^{2}}}.
\ee
The leading term is in agreement with (\ref{M-AZ})
\bea
\label{MaAZexp}
 K(z_a,z)&=&
 k_4z^4+k_6z^6+{\cal O}\left( {z}^{8}  \right),
\eea
where
$
k_4= 1/z_h^4+2\,c/z_h^2+{\cal O}
 \left( 1 \right)
$,
$k_6=
-2c/z_h^4+{\cal O}
 \left( 1 \right)$.

\begin{figure}[h!]
    \centering
 \includegraphics[width=6cm]{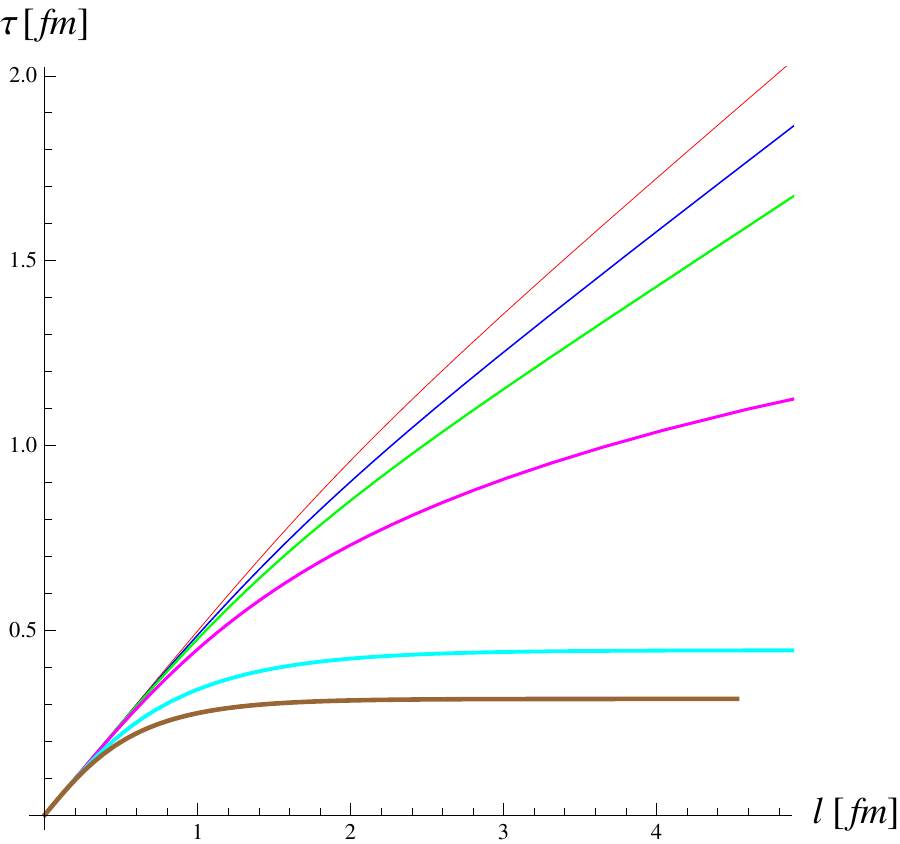}
 $A.\,\,\,$
 $\,\,\,\,\,\,\,\,\,\,\,$ 
  \includegraphics[width=6cm]{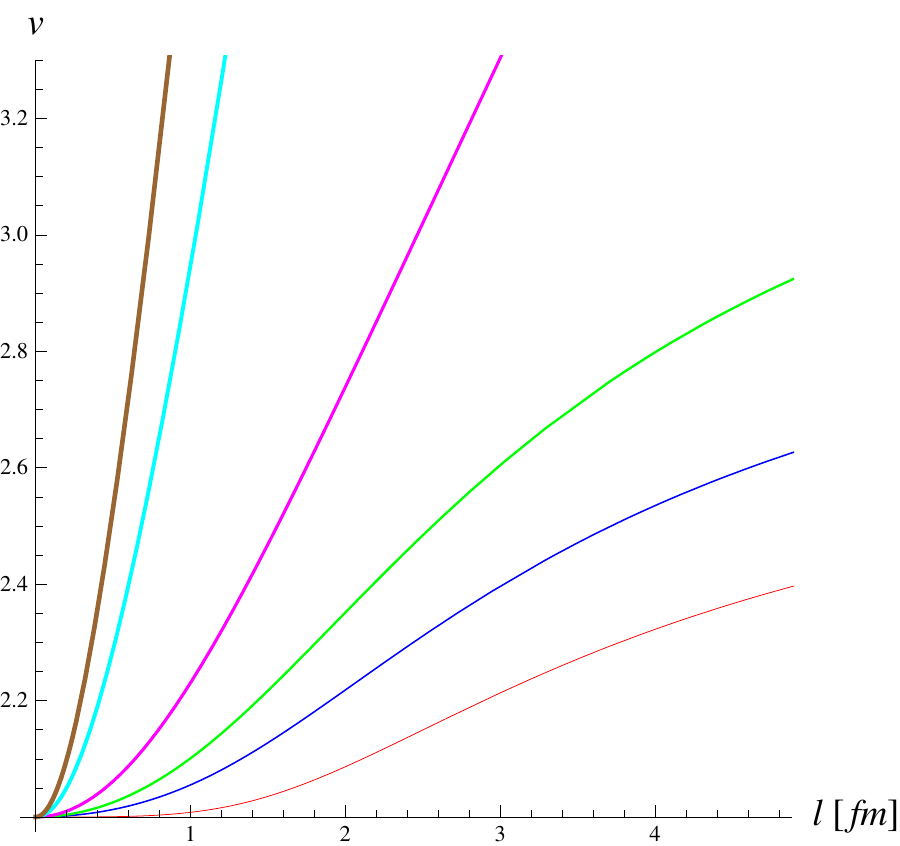}
   $B.\,\,\,$
    \caption{A. Dependencies of $\tau$ on $\ell$   for 5-dimensional metric
    (\ref{ds-f-b}) with  the confining $b$-factor (\ref{AZ}) for $c=0$ (red), $c=0.1$ (blue), $c=0.2$ (green), $c=0.5$ (magenta), $c=2.56$ (cyan) ,  $c=5.16$ (brown ) and the blackening factor (\ref{Ma}) with $z_h=1$.
B.  Dependencies of  $v=\ell/\tau$ on $\ell$ for the same parameters as in the left panel.
}
 \label{v-l-exp-m}
\end{figure}

In Fig.\ref{v-l-exp-m}.A the dependence of the  thermalization time $\tau$ on the scale $\ell$  for the metric (\ref{AZ})
 for different values of $c$ is shown. From this plot we see that increasing the confinement potential,
in fact when $c$ in formula (\ref{AZ}) increases,  we decrease the thermalization time.

In Fig.\ref{v-l-exp-m}.B. the dependence of  $v=\ell/\tau$ on $l$ for the same parameters as in  Fig.\ref{v-l-exp-m}.A is presented. We can interpret
$v=l/t$ as a propagation velocity of the thermalization. We see that the velocity increases with increasing of the factor $c$ in (\ref{AZ}).

\begin{figure}[h!]
    \centering
 \includegraphics[width=6cm]{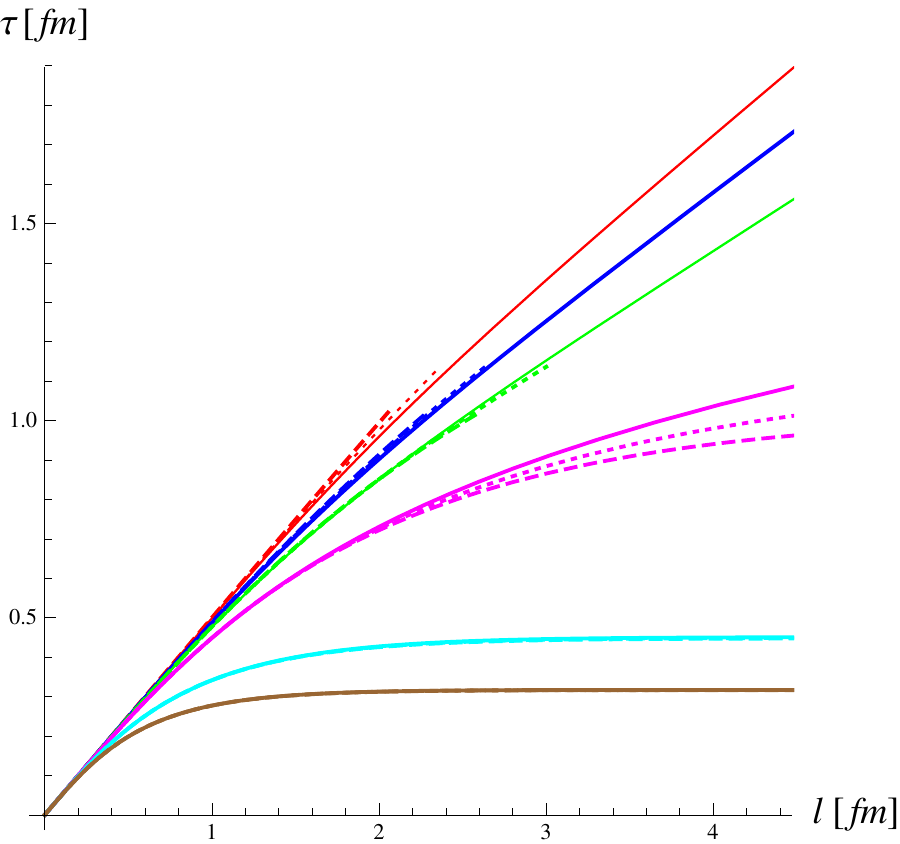}
 $A\,\,\,$
 \includegraphics[width=3.4cm]{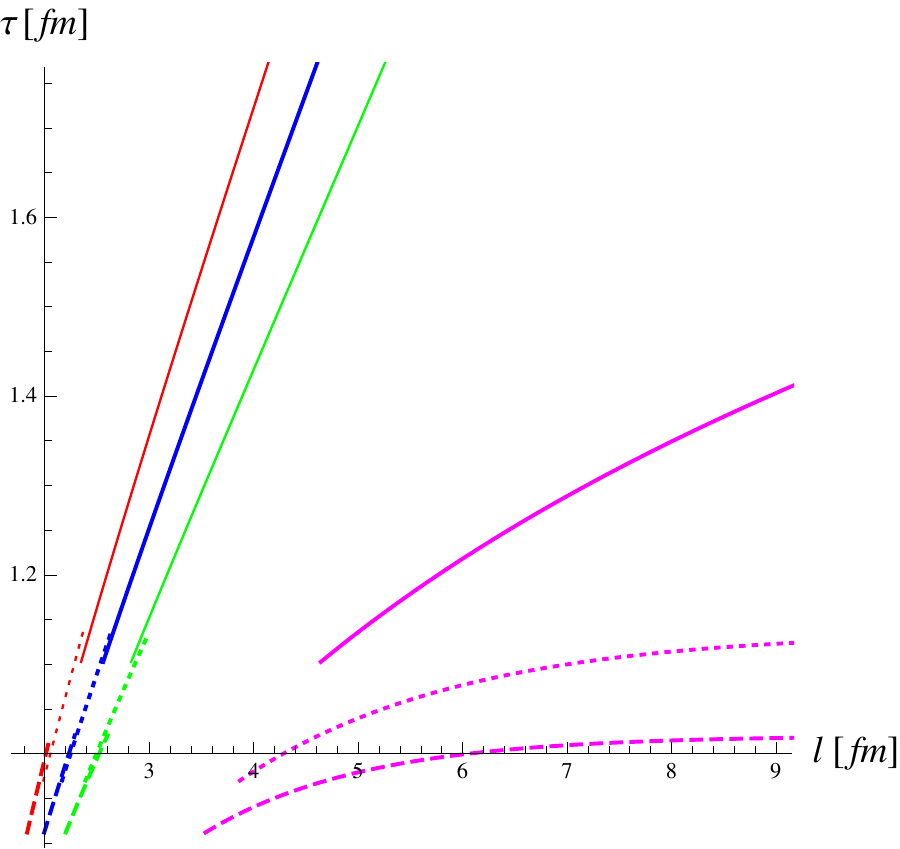}
 $B$ \includegraphics[width=3.4cm]{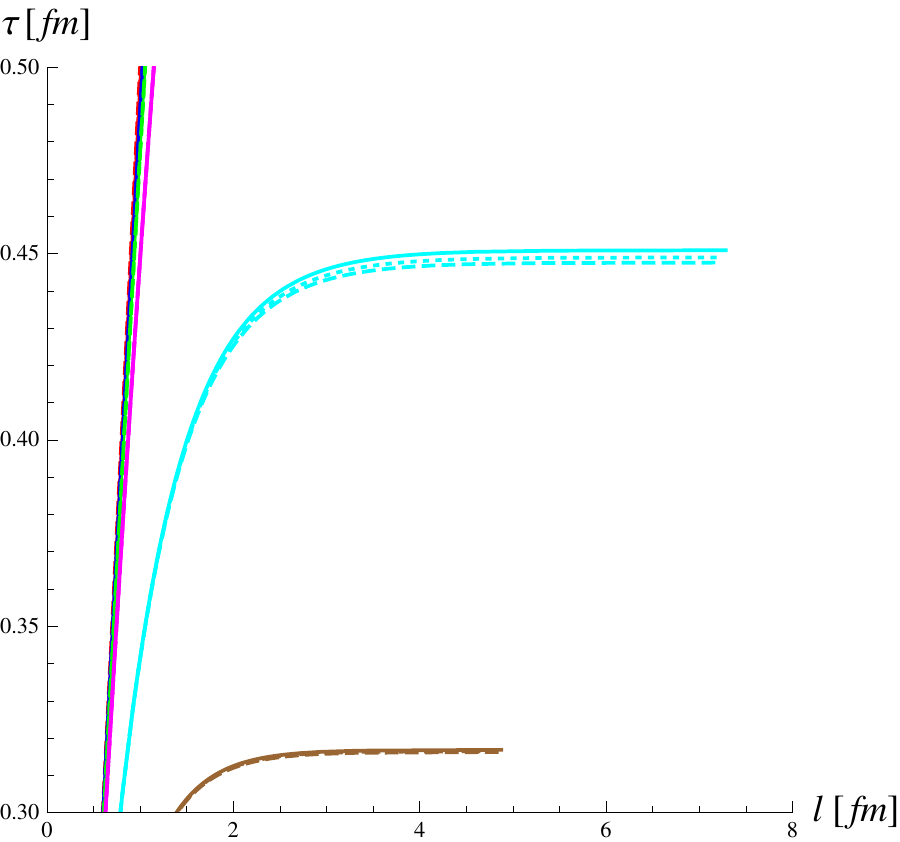}
 $C$
    \caption{A.
  Dependencies of $\tau$ on $\ell$    for the 5-dimensional metric
    (\ref{ds-f-b}) with  the confining $b$-factor (\ref{AZ})  for $c=0$ (red), $c=0.1\,fm^{-2}$ (blue), $c=0.2\,fm^{-2}$ (green), $c=0.5\,fm^{-2}$ (magenta), $c=2.56\,fm^{-2}$ (cyan), $c=5.16\,fm^{-2}$ (brown )
    and  $z_h=1fm$ (solid lines), $z_h=1.2fm$
  (dotted lines), $z_h=1.8\,fm$ (dashed lines).
B.  The zoom of the plot A  in the region  $2\,fm<\ell<9\,fm$ and $1\,fm<\tau<1.8\,fm$.
C.  The zoom of the plot A in the region $0<\ell<8\,fm$ and $0.3\,fm<\tau<0.5fm$. }
 \label{v-l-exp-za-zb}
\end{figure}

In Fig.\ref{v-l-exp-za-zb}.A.  dependencies of  $\tau $ on $\ell$ for different masses of the shell and  the same parameters $c$ as in Fig.\ref{v-l-exp-m}.A  are presented. We see
that for chosen parameters the dependence on $z_h$ is very small.
We see that increasing $z_h$ for $AdS$ case we increase the thermalization time.
The same is true for $c=0.1$. However,
when we increase  $c$ so that $0.2<c<0.5$   the dependence on $z_h$ becomes more essential (the distances between the magenta lines
are larger as compared with distances of green lines) and when increasing $z_h$
we decrease the thermalization time. We also see, Fig.\ref{v-l-exp-za-zb}.B.,
that in considered cases only up to some given distance the thermalization is possible. This is related with the breaking of geodesics with two large endpoints distance in the confining background.

\begin{figure}[h!]
    \centering
\includegraphics[width=5cm]{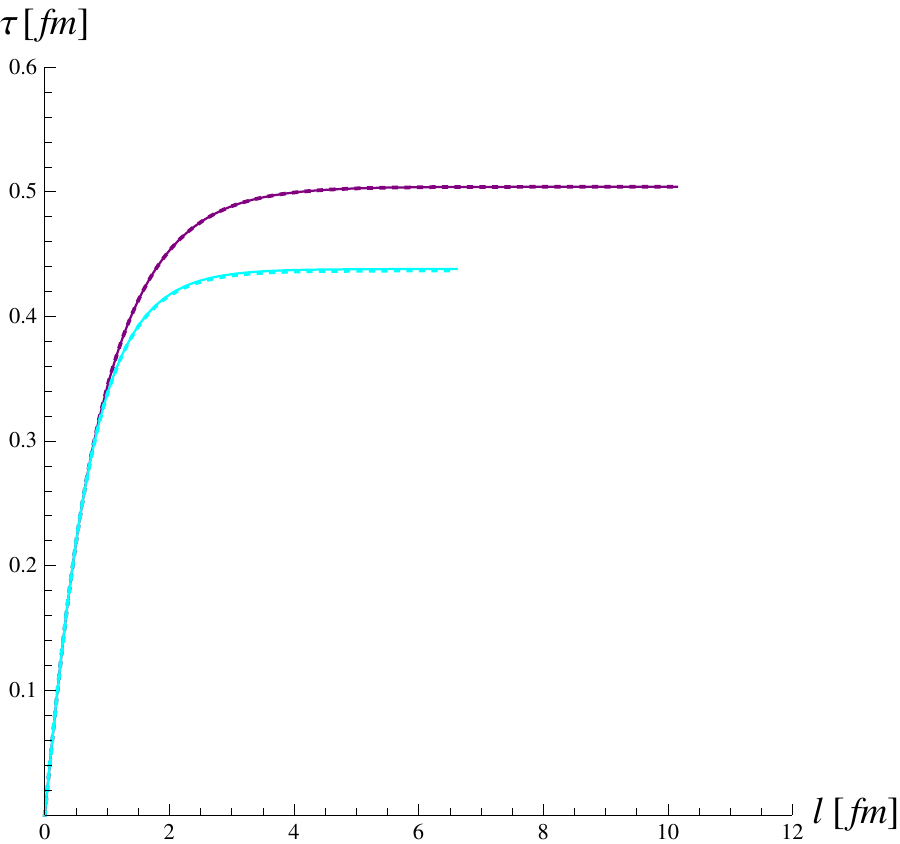}$\,\,A\,\,\,\,$
\includegraphics[width=4cm]{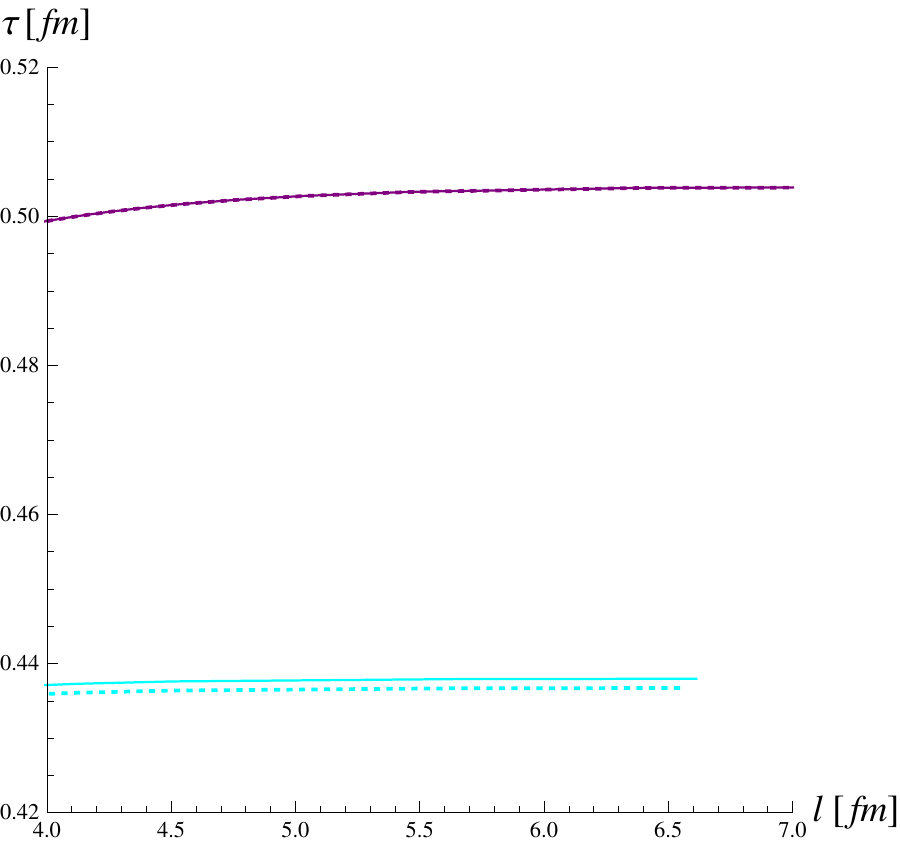}
$B\,$ \includegraphics[width=4cm]{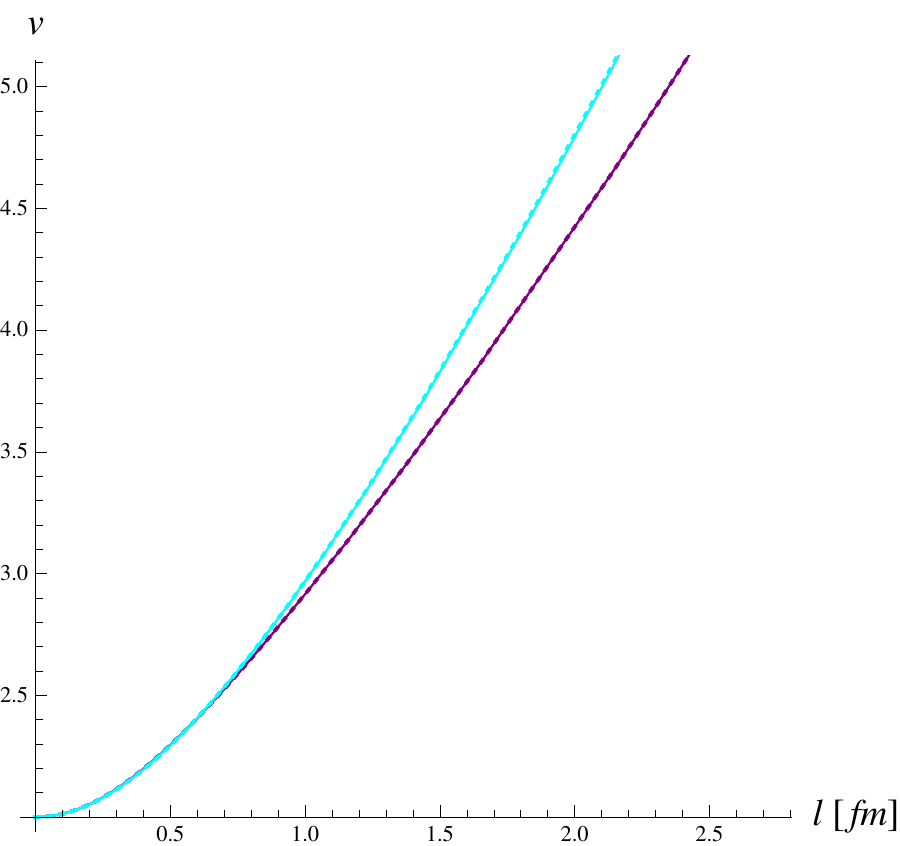}
 $C$
    \caption{A.
    Dependencies of $\tau$ on $\ell$    for the 5-dimensional metric
    (\ref{ds-f-b}) with  the confining $b$-factor (\ref{AZ})  with $c=2.56$
    and different blackening functions: blackening function (\ref{f-a})
    for  $z_h=1.2$ and $z_h=1.8$ corresponds to solid and dashed blue lines;  blackening function (\ref{MaAZ})
    for  $z_h=1.2$ and $z_h=1.8$ corresponds to solid and dashed red lines.
    B.  The zoom of the plot A. C. Dependencies of $\tau/\ell$ on $\ell$  for the same parameters as in the plot A.
}
 \label{c256}
\end{figure}

$$\,$$

In Fig.\ref{c256} the thermalization process is presented for more realistic values of  $c$, namely $c=2.56\,fm^{-2}$.   Dependencies of $\tau$ on $\ell$    for 5-dimensional metric
    (\ref{ds-f-b}) with  the confining $b$-factor (\ref{AZ})  with $c=2.56$
    and different blackening functions: blackening function (\ref{f-a})
    for  $z_h=1.2$ and $z_h=1.8$ corresponds to solid and dashed blues lines;  blackening function (\ref{MaAZ})
    for  $z_h=1.2$ and $z_h=1.8$ corresponds to solid and dashed red lines. We see that corrections (\ref{MaAZexp}) do not play an essential role for small distances, but at large distances the blackening factor (\ref{M-AZ}) admits longer geodesics as compared with (\ref{MaAZ}).

\subsubsection{Estimation for power metric}\label{S:2.2.3}

It is instructive to compare  estimations obtained in the previous  subsection
\ref{S:2.2.2} with results obtained for the intermediate
metric \cite{Ageev:2014mma}
\be\label{a-b}
ds^2_{inter}=\left(\frac{L_{eff}}{z}\right)^{2a}\left(-dt^2+dz^2+dx^2\right),\ee
 In this case the blackening function is
\be
\label{f-a}
f(z)=1-\frac{z^{da+1}}{z_h^{da+1}},\ee
and the thermalization time $\tau$ at the scale $l$ is
\bea\label{x-d}
\ell &=&2s\int_0^{1}\frac{w^{a}dw}{\sqrt { (1-\frac{s^{1+da}}{z_h^{1+da}}w^{1+da})\left(1-w^{2a}\right)}},\\\label{t-d}
\tau&=&
s\int _{0}^{1}\frac{dw}{\,(1-\frac{s^{1+da}}{z_h^{1+da}}w^{1+da})}.
\eea

\begin{figure}[h!]
    \centering
     \includegraphics[width=5cm]{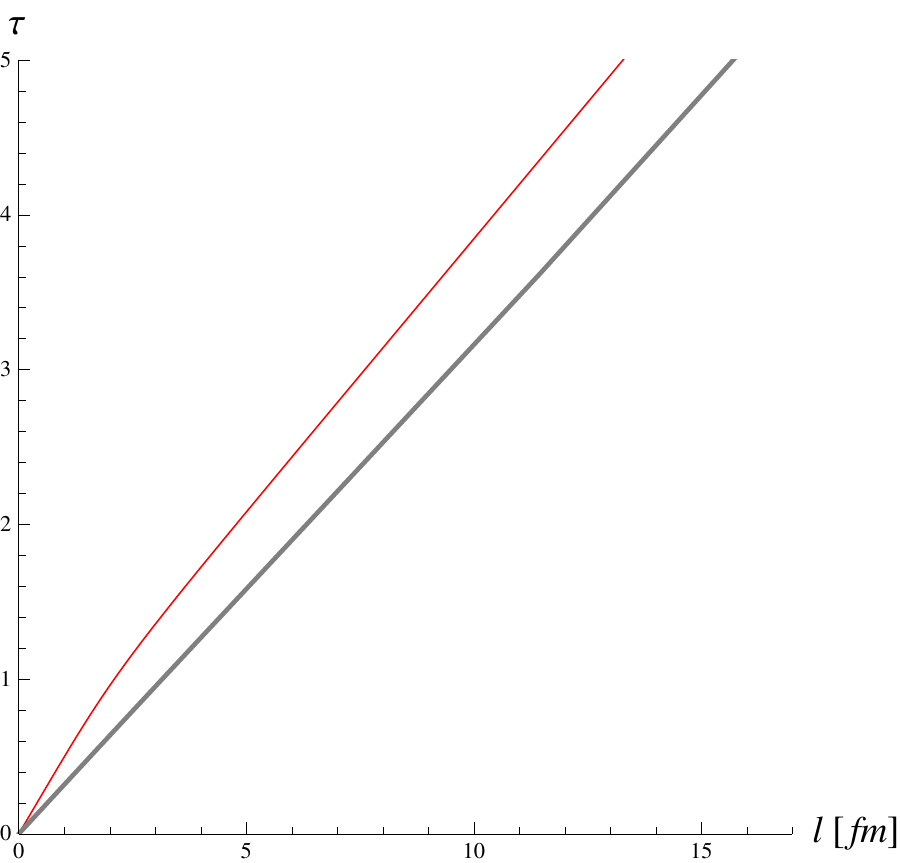}$\,\,A\,\,\,\,\,\,\,\,\,$
    \includegraphics[width=4cm]{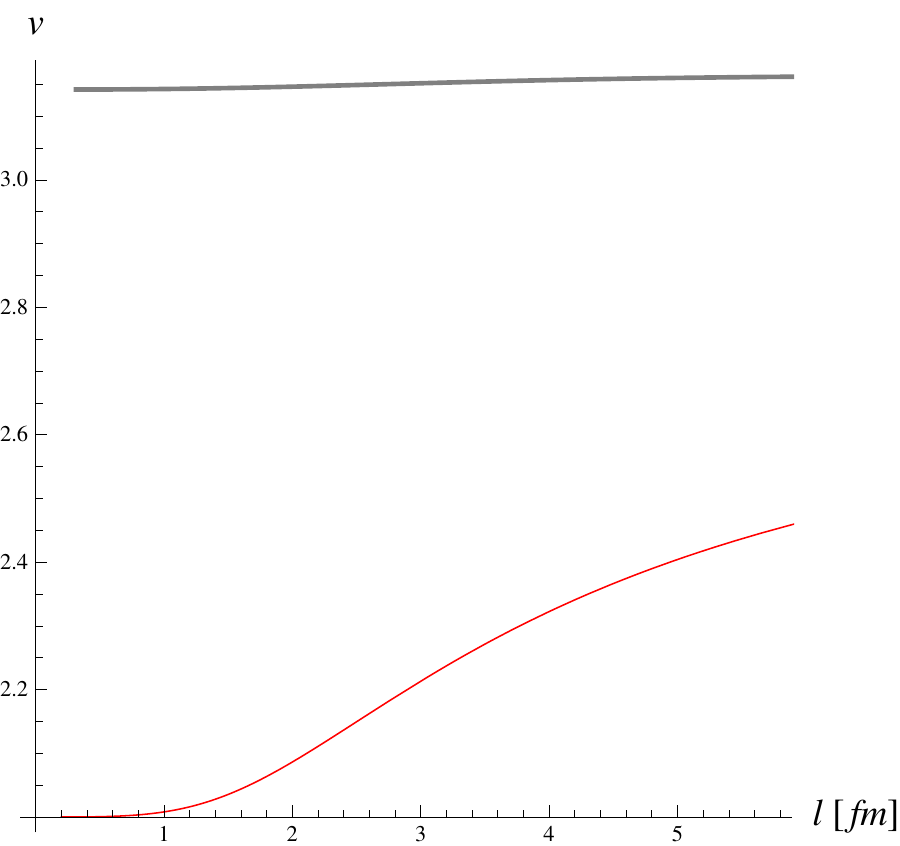}$\,\,\,B$
  \caption{ A. Dependencies of $\tau$ on $\ell$    for 5-dimensional metric
    (\ref{ds-f-b}) with  the power-law $b$-factor (\ref{a-b})  for  $a=1$ (solid red line), i.e. the $AdS_5$ case and
 $a=0.5$ (thick  gray line). B. Velocities for the same parameters.  }
 \label{a05}
\end{figure}

In Fig.\ref{a05}.A  dependencies of $\tau$ on $\ell$    for 5-dimensional metric
    (\ref{ds-f-b}) with  the power $b$-factors (\ref{a-b}):   $a=1$ (solid line), i.e. the $AdS_5$ case, and
 $a=0.5$  (dashed   line) are presented. We see that  decreasing $a$
 we decrease the thermalization time. From Fig.\ref{a05} we see that the velocity of propagation of thermalization increases with decreasing $a$.

It is interesting to note that the dependence of $t$ on $l$ does not depend on the position of the horizon for $a=1$, i.e.
\be
\label{tau-1}
\tau(z_h,l)|_{a=1}=\tau(1,l)|_{a=1}\ee

\begin{figure}[h!]
    \centering
     \includegraphics[width=7cm]{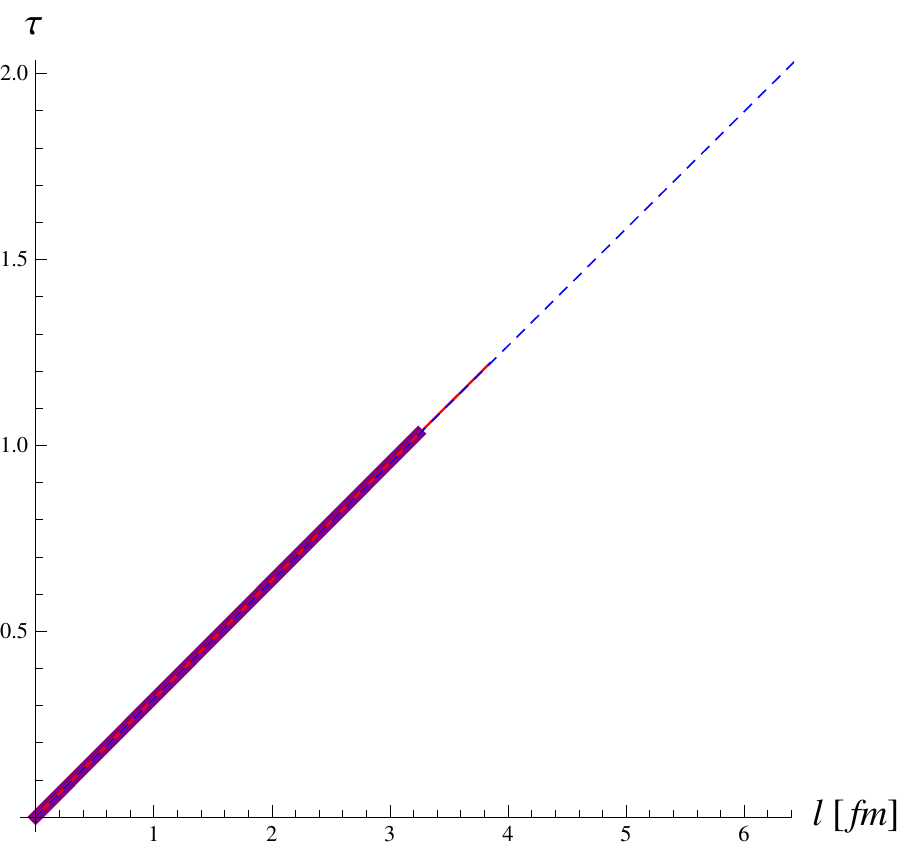}
 \caption{ A. Dependencies of $\tau$ on $\ell$    for 5-dimensional metric
    (\ref{ds-f-b}) with  the power $b$-factor (\ref{a-b})  for
 $a=0.5$ and blackening factor (\ref{f-a}) with different $z_h$: $z_h=1$ (blue dashed line), $z_h=1.5$ (red solid line), $z_h=2.5$ (purple thick solid line).  }
 \label{a05-zh}
\end{figure}
The confirmation of (\ref{tau-1}) is presented  in Fig.\ref{a05-zh}. Indeed, we see in the plots in Fig.\ref{a05-zh} that three lines, the purple thick line, the red solid  line and the dashed blue one coincide, i.e. the thermalization time does not depend on $z_h$.

From (\ref{x-d}) and (\ref{t-d}) it is evident that
\bea\label{X-d}
\frac{\ell}{z_h}&=&2S\int_0^{1}\frac{w^{a}dw}{\sqrt { (1-S^{1+da}w^{1+da})\left(1-w^{2a}\right)}},\\\label{T-d}
\frac{\tau}{z_h}&=&
S\int _{0}^{1}\frac{dw}{\,(1-S^{1+da}w^{1+da})},
\eea
here we introduce a new  parameter $S=s/z_h$. The system of equations (\ref{X-d})
and (\ref{T-d})  is nothing but  the system of equations that defines the thermalization time in units of $z_h$ for the Vaidya model with  the unique mass, i.e.
\be
\tau=z_h\,\tau_{therm} (1,\frac{\ell}{z_h}).\ee

\begin{figure}[h!]
    \centering
  \includegraphics[width=6cm]{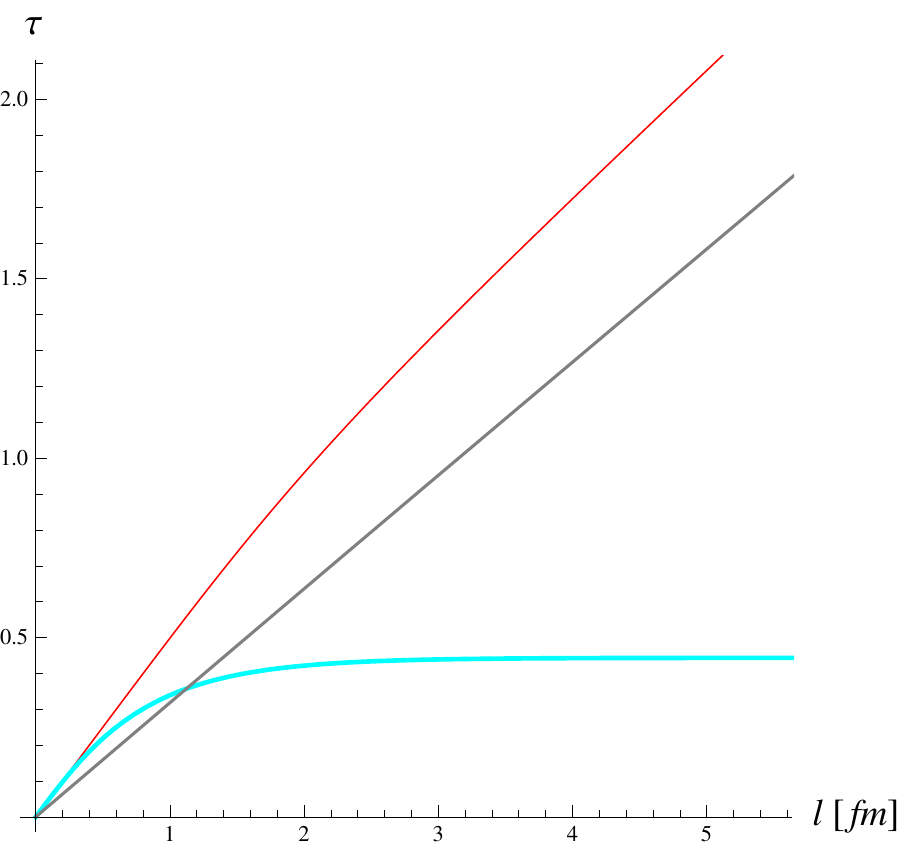}$\,\,A$\,\,\,\,\,\,\,\,\,\,\,
 \includegraphics[width=6cm]{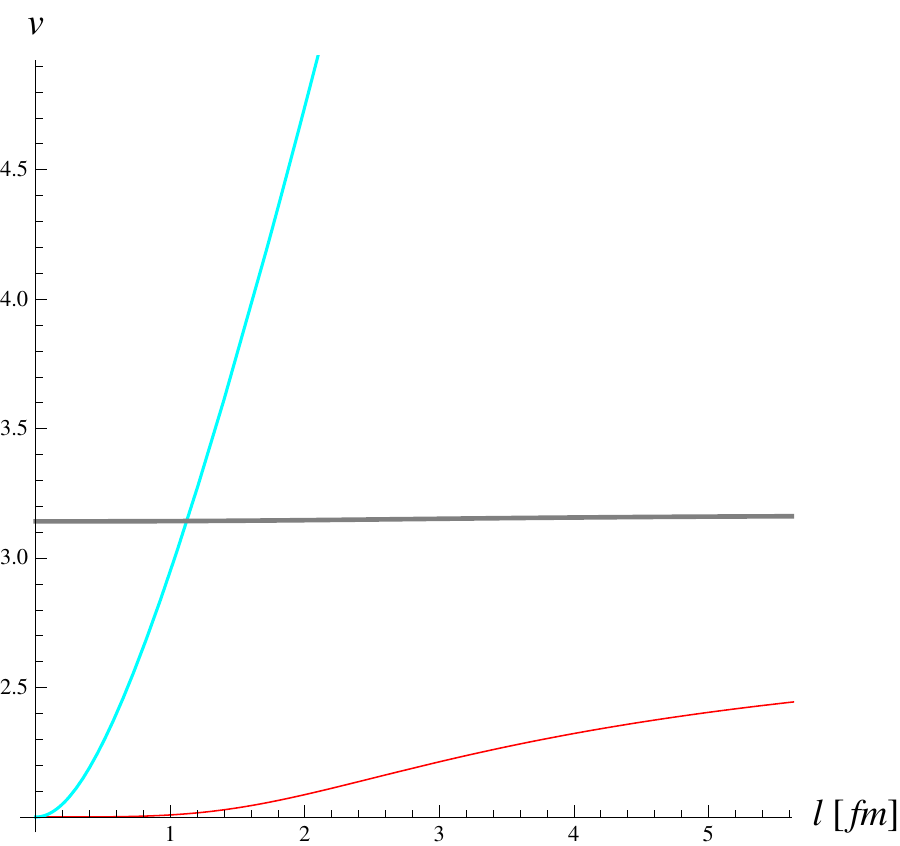}$\,\,\,B$
    \caption{ A. Dependencies of $\tau$ on $\ell$    for 5-dimensional metric
    (\ref{ds-f-b}) with  different $b$-factors:  the     thin red line corresponds to power factor (\ref{a-b}) with $a=1$ (AdS); the
  gray  line corresponds  to $a=0.5$   and the thick cyan  line corresponds to  the factor (\ref{AZ}) with $c=2.56$. B.
Dependencies of velocities of the  propagation of thermalization
on the characteristic distance.}
 \label{a1-a05-conf}
\end{figure}

Note that in \cite{Ageev:2014mma} we have shown that the b-factor for the intermediate metric
(\ref{a-b}) with $a=0.5$ and   $L_{eff}=20.86$ approximates the confining
b-factor for $1.2 fm<z<1.8 fm$. For $z\sim 1.1 \,fm $ the thermalization times for these two models are very close, however, for the distance  $z\sim 2 \,fm $ thermalization times are different up to factor 0.3. For larger
 distances  these two models show essentially different behaviour:
the thermalization time for the confining metric does not depend essentially on the distances, meanwhile  the intermediate model  has approximately a linear dependence of the thermalization time on the distance.

In Fig.\ref{a1-a05-conf}. A. we plot the dependence of the thermalization time for 3 different models: the AdS case (thin solid red line), the intermediate case  \cite{Ageev:2014mma} (gray  line)
and for the confining factor \cite{Andreev:2006ct} (thick cyan line).
For all cases $z_h=1\, fm$.

 The small dependence of the thermalization time on the value of the mass
  of the shell for special models  can be seen from the following considerations. Assuming that $b(z)$ satisfies  the scaling
 \be
 \label{b-k}
 b(kz)=k^{\varsigma_b}  b(z)+...,\,\,\,\,\,\, 0<z<z_h,\ee
one can reduce the problem of finding the thermalization time at the scale
 $\ell$ by the Vaidya metric with mass $M_h$  to the same problem with the unite mass. Indeed, from (\ref{b-k}) and (\ref{Ma}) the scaling for the mass $M(z)$ is
 \be
 \label{Msc}
 M^{-1}(\frac{z}{k})\approx k^{d\varsigma_b-1}M^{-1}(z).
 \ee
 Performing the rescaling
 \be
 z=k_a\tilde z,\,\,\,\,x=k_a\tilde x,\,\,\,\,t=k_a\tilde t,\ee
 with
 \be k_a=M_a^{\frac{1}{d\varsigma_b-1}}\ee
 we recast the metric (\ref{ds-f-b}) to the same form in terms of the tilde-coordinates and  the blackening function with a unit mass  $
f(1, \tilde z, \tilde v)$,
 \bea
\label{ds-nu-g1}
ds^2&=&M_a^{\frac{2\varsigma_b+2}{d\varsigma_b-1}}b^2(\tilde z)
\,\left(-f(1, \tilde z,\tilde v)\,d\tilde v^2-2d\tilde vd\tilde z+d\vec{\tilde x}^2\right).
\eea

Therefore,
\be
\tau_{therm} (z_a,l) =k_a\tilde \tau_{therm}(1, \tilde \ell)=k_a \tau _{therm}(1, \ell/k_a).\ee
 Assuming that
 \be
 \tau_{therm}(1,\ell)=C_1\ell+C_2\ell^2+...\ee
 we get
 \be
\tau_{therm}(z_a,z_b,\ell)\approx C_1 +C_2\ell^2/k_a^2.\ee

  \subsubsection{Estimation of the formation time of a trapped surface  by non-local correlation functions}
We can also estimate the  trapped surface formation  time  by a characteristic size of the trapped surface \cite{Ageev:2014mma}, i.e.
\be
\label{timetherm}
\tau_{therm}\sim \frac{z_b-z_a}{v_z}.\ee
Here $v_z$ is the velocity of propagation of a signal along the $z$-direction. We can estimate the velocity of propagation of the signal along the $z$ direction in two ways. We can relate  two points on the boundary by the geodesic or the string worldsheet
stretched   on the static quarks world lines located at these points.
 The geodesic as well as the world\-sheet have  maximum   z-coordinates, $z_*$. Varying the end points we vary $z_*$ and we can estimate
 \be
 \label{z-Deltaz}
 v=\frac{\Delta z_*}{\Delta x}\cdot c,\ee
here we take into account that any propagation in the $x$-direction is limited by the light velocity. Below we assume $c=1$.

{\it Estimation with string}.  The relation between the interquark distance  $x$  and
the string maximum holographic coordinate $z_{m}$  is given by
\be
\label{x-zm}
x=2\int_{0}^{z_m} \frac{dz}{\sqrt{\frac{b^4(z)}{b^4(z_m)}-1}}.
\ee
From this formula we have
\be
\label{vz}
v=\frac{\Delta z}{\Delta x}=\sqrt{\frac{b^4(z)}{b^4(z_m)}-1}.\ee
We see that (\ref{vz}) defines the z-dependent velocity. In particular, considering this estimation in the confining background \cite{Ageev:2014mma,AI-proc}  where  $1.2 \,fm<z_m<1.8 \,fm$ and we get  $\Delta x=\Delta z_m/2.4$.  This estimation   gives the trapped surface formation time $\tau_{therm}\approx 0.25\, fm$.

{\it The estimation with geodesics}  in the intermediate background  gives 
$
\Delta x=\frac{\pi}{2} \Delta  z_m
$, which is 4 times   longer as compared with the string estimation.

\section{Thermalization in anisotropic backgrounds}
\subsection{Setup}
\subsubsection{Anisotropic metrics}
 As an  anisotropic background (with the space anisotropy) we  consider  a five-dimensional Lifshitz-like metric \cite{Taylor:2008tg,Pal,Bueno:2012sd}
\be\label{L-as}
ds_{Ll}^{2} =L^{2}\left[\frac{\left(-dt^{2} + dx^{2}\right)}{z^{2}} + \frac{\left(dy^{2}_{1} +dy^{2}_{2}\right)}{z^{2/\nu}} +  \frac{d z^{2}}{z^{2}}\right].
\ee
Let us remind that the anisotropic Lifshitz metric with space symmetry is given by \cite{0808.1725,Taylor:2008tg,Tarrio:2011de,Gouteraux:2011ce,Huijse:2011ef,Dong:2012se,Bueno:2012sd,1212.2625}
\be\label{L-s}
ds_{Lif}^{2} =L^{2}\left[\frac{-dt^{2} }{z^{2}} + \frac{\left(dx^{2}+dy^{2}_{1} +dy^{2}_{2}\right)}{z^{2/\nu}} +  \frac{d z^{2}}{z^{2}}\right].
\ee
Both metrics for $\nu = 1$ are reduced  to the Poincare patch of $AdS_5$.

\subsubsection{Thermalization due to shock waves collision}
This scenario  assumes that
the main part of multiplicity is produced in an anisotropic regime  and this part of multiplicity
can be estimated by the trapped surface  produced under a collision of the two shock waves in an anisotropic background. This scenario is accepted in the recent paper \cite{Aref'eva:2014AG}, where collisions of shock waves in the Lifshitz-like background  have been considered.

As a model of anisotropic background we consider  a five-dimensional Lifshitz-like metric
(\ref{L-as}).
The shock domain  wall moving in the $v$-direction is given by deformation of the metric (\ref{L-as})
\be
ds_{DW,Ll}^{2} =ds_{Ll}^{2}+ L^{2}\frac{\phi(z)\delta(u) }{z^{2}}du^{2},
\ee
with the profile function $\phi(z)$ satisfying
\be\label{4.2.1b}
\frac{\partial^{2} \phi(z)}{\partial z^{2}}  -  \left(1 + \frac{2}{\nu}\right)\frac{1}{z} \frac{\partial \phi(z)}{\partial z} = -16\pi G_{5} J_{uu},\,\,\,\,\,\,
J_{uu} = E \left(\frac{z}{L}\right)^{1+2/\nu}\delta(z  - z_{*}),
\ee
where $z_*$ is the $z$-coordinate of the collision point.
The entropy can be written in terms of  $z_{a}$ and $z_{b}$ defining the location of the trapped surface
\be\label{4.3.6a}
s = \frac{\nu}{4G_{5}}\left(\frac{1}{(z_{a})^{2/\nu}} - \frac{1}{(z_{b})^{2/\nu}}\right).\ee
The analog of relation (\ref{a-b-E}) is
\be
 \label{a-b-E-nu}
 z_a^{-1-2/\nu}+z_b^{-1-2/\nu}=\frac{8\pi G_5 E}{L^{2/\nu+3}}.
 \ee
The maximal entropy is achieved  at $z_b\to \infty$ and it is
\be
s = \frac{\nu}{4G_{5}}(8\pi G_5)^{2/(\nu+2)}E^{2/(\nu+2)}.\ee

The leading asymptotic  gives  rise to the value of multiplicity, which is the most compatible to the experimental data, for $\nu =4$ \cite{Aref'eva:2014AG}.
Note, that as in the isotropic case for intermediate values of the energy  we have
to take into account the next to leading term, but now we do not have a restriction on $z_a$ from below since we do not obliged to fit our metric to the metric with a given $b$-factor. We have changed the background on which we consider the collisions of domain walls. It would be interesting to find "top-down" motivations for consideration the
anisotropic background  as a holographic model  for HIC.

\subsection{Thermalization time in the anisotropic
background}

 \subsubsection{Estimation with Vaidya metric}
We can  estimate the thermalization time in the anisotropic
background
\bea
\label{ds-nu-g-an}
ds^2&=&b^2(z)\left(-\frac{f(z_h,z)}{z^{2(\nu-1)}}dv^2-
2\frac{dvdz}{z^{\nu-1}}+d\vec{x}^2\right),\\
\label{f-K}
f(z_h, z) &=& 1-\frac{K(z)}{K(z_a)}.\eea

For metric (\ref{ds-nu-g-an}) with $f(z_h, z)$ as in (\ref{f-K})
we have

\bea
\ell&=&
 2s\int_0^{1}\frac{b(s)}{b(sw)}\,\frac{dw}{\sqrt {\left(1-\frac{K(sw)}{K(z_a)}\right) \cdot\left(1-\frac{b^2(s)}{b^{2}(sw)}\right)}},
 \\
 \tau&=&
s^\nu \int _{0}^{1}\frac{dw}{w^{1-\nu}\,\left(1-\frac{K(sw)}{K(z_a)}\right)}.\eea

\begin{figure}[h!]
    \centering
 \includegraphics[width=5cm]{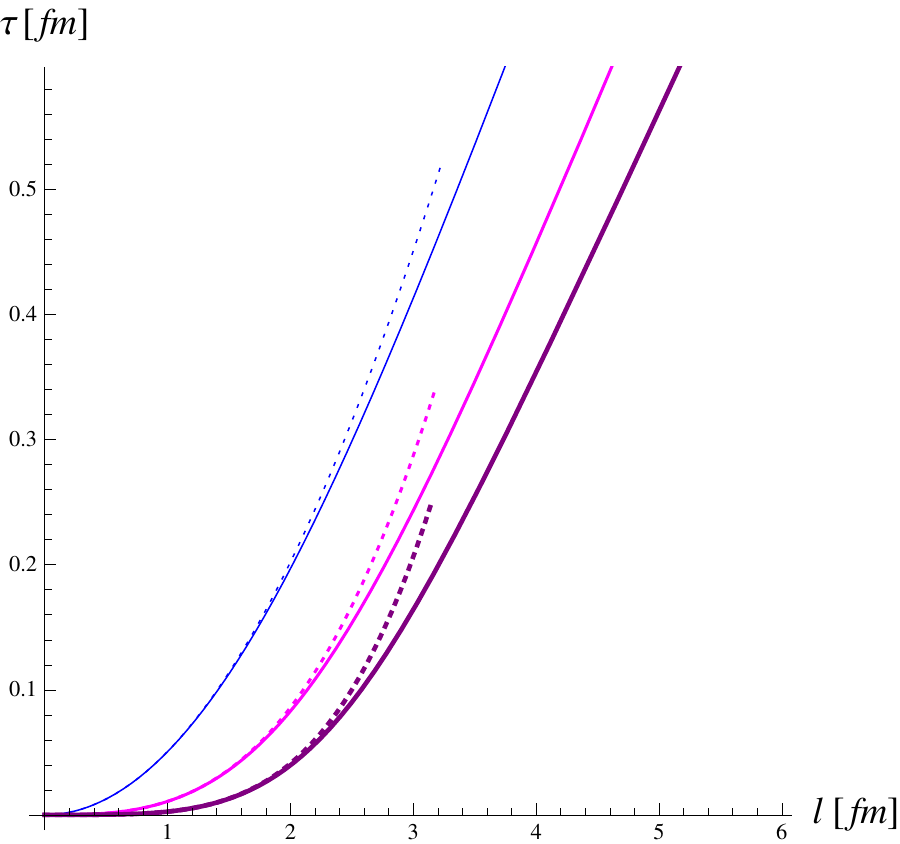}$A\,\,\,$
 \includegraphics[width=3.5cm]{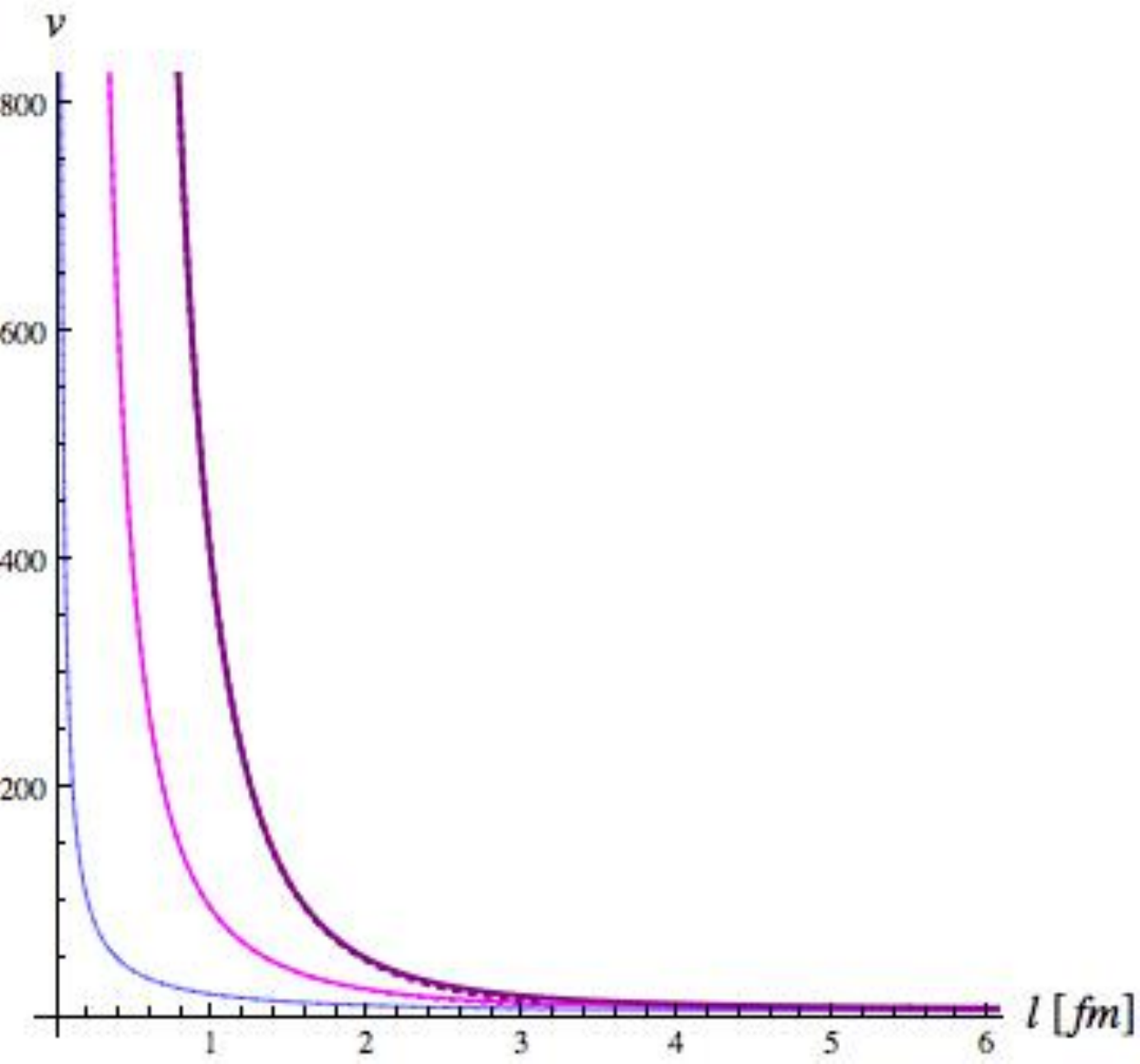}$B\,\,\,\,\,\,\,\,\,\,\,\,$
 \includegraphics[width=3cm]{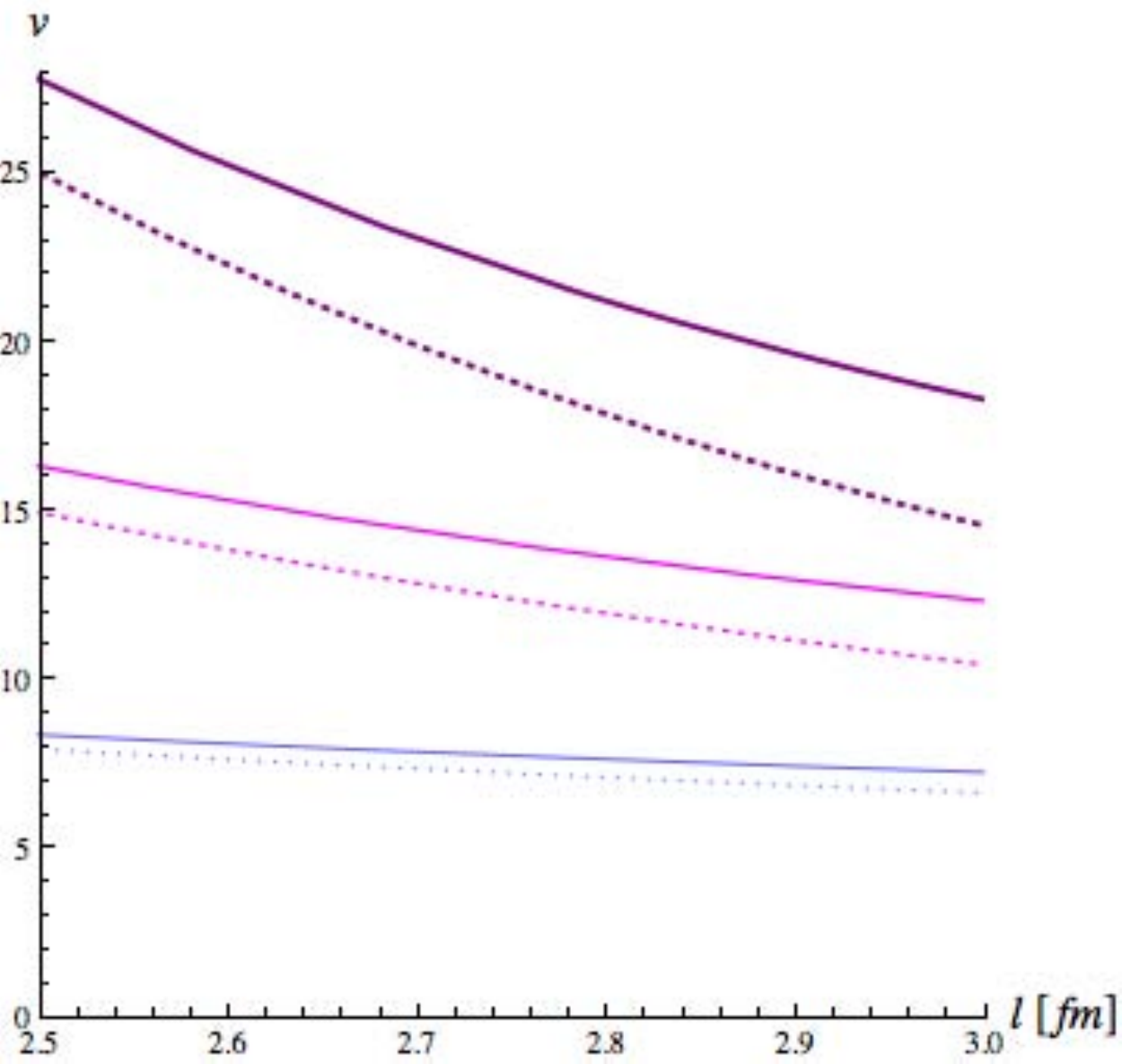} $\,C$
\caption{ A. Dependencies of $t$ on $l$  for 5-dimensional case
  for $a=0.5$  and different $\nu$  and different $z_h$:
  solid lines $z_h=1$ and dotted lines $z_h=2$. Blue lines: $\nu=2$;
  magenta lines: $\nu=3$; purple lines: $\nu=4$. B. Dependencies of velocity of thermalization propagation
on the distance. The zoom to the plot B.  }
 \label{Aniz-zh}
\end{figure}

For the power-law $b$-factor, $b(z)=(L/z)^a$
the blackening factor is given by (\ref{f-K}) with
\be
\label{K-an}
K(z)=z^{ad+\nu}\ee
Thermalization for these models have been considered in \cite{Keranen:2011xs,alishahiha:2012}. In Fig.\ref{Aniz-zh}
we show the dependence of the thermalization time on the distance for the  model
with $a=0.5$ and different $z_h$. We see that increasing the anisotropy  the dependence on the horizon position increases as well. We also see that the breaking of the geodesics appears at smaller distances for the case of  small $z_h$ (heavy black masses). 

For the confining $b$-factor by the analogy with (\ref{M-AZ}) we use the
"phenomenological" blackening function
\be
\label{f-an-AZ}
K(z) = z^{3+\nu}.\ee
In Fig.\ref{an-AZ} the influence of the anisotropy on the thermalization time for the confining metric (\ref{ds-nu-g-an}) for different $c$ is presented.  We see that anisotropy essentially decreases the thermalization time. Meanwhile there is no essential dependence of the thermalization time on the mass of the shell.

\begin{figure}[h!]
    \centering
 \includegraphics[width=10cm]{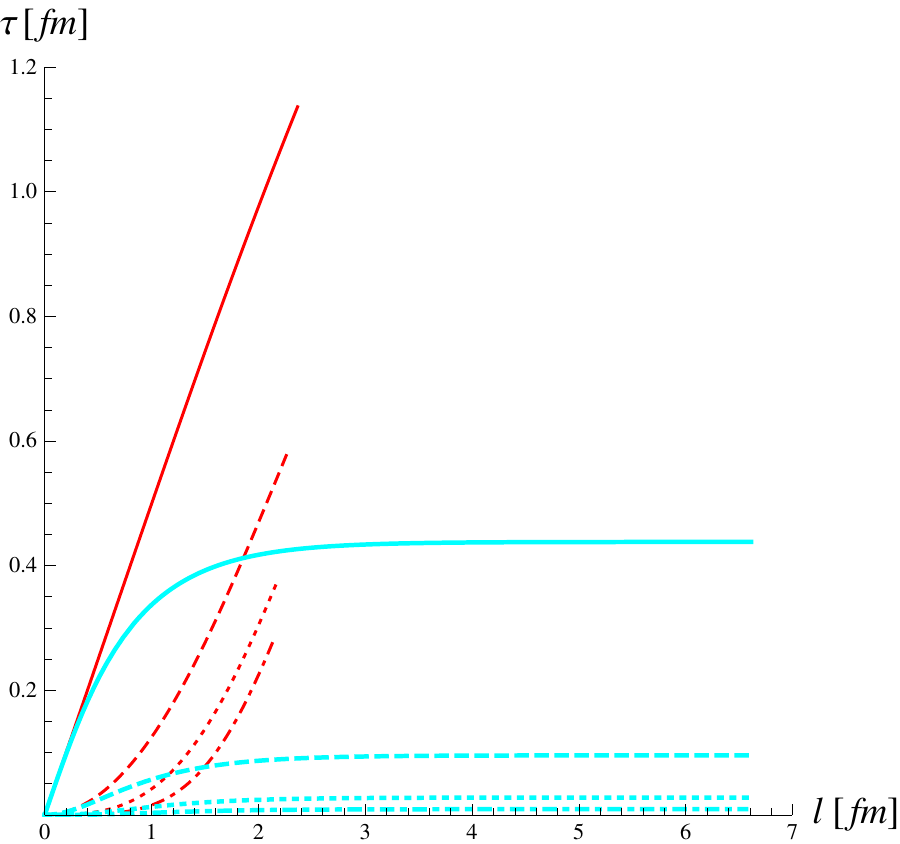}
  \caption{ Dependencies of $\tau$ on $\ell$    for the 5-dimensional metric
    (\ref{ds-nu-g-an}) and the blackening factor (\ref{f-an-AZ}) for
    different values of $c$ and  $\nu$.
Red lines   correspond to the  AdS case, $c=0$ and cyan lines to the confining b-factor with $c=2.56 fm^{-2}$.
The isotropic case, $\nu=1$, is shown by  solid lines,  anisotropic cases with $\nu=2$ are shown by dashed lines,  $\nu=3$ by dotted lines
  and $\nu=4$ by dot-dashed lines. For all cases $z_a=1.2 fm $.
 }
 \label{an-AZ}
\end{figure}

\begin{figure}[h!]
    \centering
\includegraphics[width=9cm]{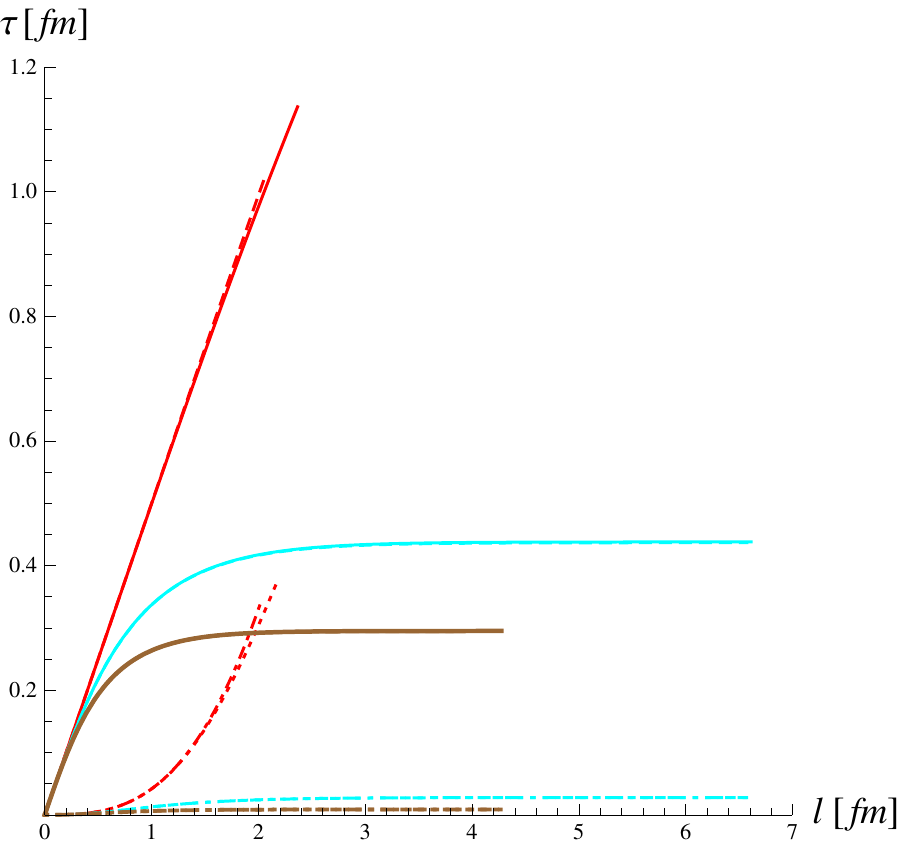}$A\,\,\,\,\,\,$
\includegraphics[width=4cm]{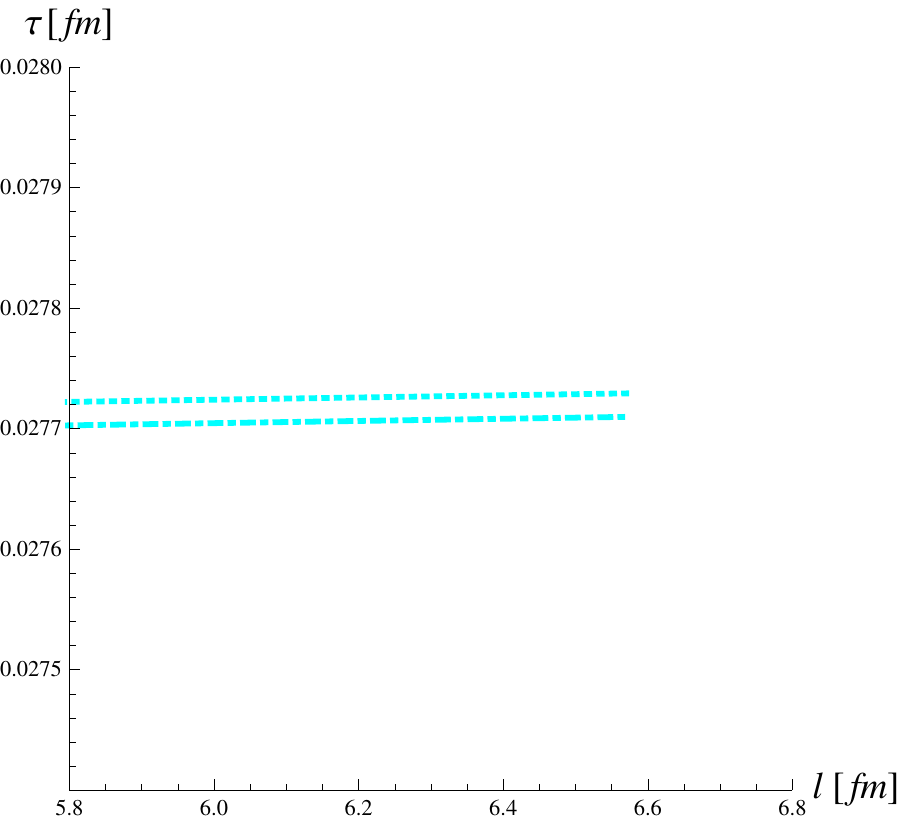}$B$
  \caption{ Dependencies of $\tau$ on $\ell$    for 5-dimentional  metric
    (\ref{ds-nu-g-an}) and blackening factor (\ref{f-an-AZ}) for
    different $c$,  different $\nu$ and different $z_a$.
Red lines   correspond to the  AdS case, $c=0$:
  isotropic case ($\nu=1$) solid $z_a=1.2$ and $z_a=1.8$ dashed  lines, anisotropic case ($\nu=3$)  $z_a=1.2$ dotted and $z_a=1.8$ dot-dashed  lines.
 Cyan lines correspond to the confining $b$-factor  with $c=2.65$:
 isotropic case ($\nu=1$) $z_a=1.2$  solid and  $z_a=1.8$  dashed lines, anisotropic case ($\nu=3$)  $z_a=1.2$   dotted and   $z_a=1.8$  dot-dashed lines.
Brown lines correspond to the confining $b$-factor with $c=5.75$ and different anisotropy
factors and different $z_a$:
 isotropic case  $z_a=1.2$  solid and  $z_a=1.8$ dashed lines, anisotropic case ($\nu=3$)  $z_a=1.2$ dotted and   $z_a=1.8$ dot-dashed lines. B. The zoom of plot A, $5.8\,fm<\ell<6.8\,fm$,
 $0.027\, fm<\tau<0.028\, fm$.
}
 \label{an-AZ}
\end{figure}

\subsubsection{Estimation with non-local correlation functions}
To estimate the trapped surface formation time we use an analog of estimations (\ref{timetherm}) and (\ref{z-Deltaz}). 
Considering the relation between the interquark distance along $y_1$-direction  and
the maximum of the string profile holographic coordinate $z_{m}$ we get the relation
\be
\label{x-zm}
y_1=2z_m^{1/\nu}\int_{0}^{1} \frac{dy}{\sqrt{1-z^{2-2/\nu}}}.
\ee
In particular for $\nu =4$ we have $y_1=2.22\,z_m^{1/4}$ and 
therefore, $\Delta y\approx0.55\,\frac{\Delta ( z_m)}{z_m^{3/4}}$, that for the trapped surface located at the intermediate zone $1.3\,fm< z<1.8\,fm $ gives $\tau_{therm}\approx0.2 \,fm/c$.

Note, that the estimation with geodesics, gives 
$
y_1=2\nu z_m^{1/\nu}
$, which is once again in four times longer as compared with  the string estimation.

One can compare these estimations of the trapped surface formation time with the thermalization time at scale about $2 \,fm\div 4 \,fm$, that is according sect. 3.2.1 for $\nu=4$  is
$\sim 0.05\,fm$.
  
\section{Conclusion}

As it has been mentioned in the Introduction,   the entropy of the  black hole produced in the
shock wave  collision predicts    multiplicities
 for heavy ion collisions at RHIC and LHC only for  special backgrounds.  In this paper we have estimated   the thermalization time
for these cases.

 In particular,   we  have estimated the anisotropic thermalization time  in
a holographic bottom-up AdS/QCD  confinement backgrounds  that
provides the Cornell potential and  QCD $\beta$-function.
We have shown that thermalization time  is up to 5 times   faster comparing to
 the isotropic case. It is  interesting  that we have not seen
essential dependence of the thermalization time on the temperature, i.e. our method predicts the same order of anisotropic thermalization time for
RHIC and LHC.

\section*{Acknowledgments}
 
This work is supported by the Russian Science Foundation (project 14-50-00005,
Steklov Mathematical Institute).
The author is grateful Dmitrii Ageev, Anastasia Golubtsova and Igor Volovich  for  useful discussions.

\end{document}